\documentclass[]{aa}

\usepackage{graphics}

\begin{document}

    \thesaurus{01         
               (08.01.1;  
                08.16.3;  
                08.09.2 HD\,140283;  
                10.01.1)  
              }
    \title{The boron absorption line at 2089.6\,\AA\ in HD\,140283 ?
           \thanks{Based on observations made with the NASA/ESA
                   {\it Hubble Space Telescope}, obtained at the
                   Space Telescope Science Institute, which is
                   operated by AURA, Inc., under contract NAS 5-26555
                  }
          }

    \author{Patrik Thor\'en and Bengt Edvardsson}

    \institute{Uppsala Astronomical Observatory, Box 515,
               SE-751\,20 Uppsala, Sweden
              }

    \authorrunning{P. Thor\'en and B. Edvardsson}
    \titlerunning{Boron @ 2089.6\,\AA\ in HD\,140283 ?}
    \offprints{Bengt Edvardsson}

    \date{Received 8 September 2000; accepted 14 November 2000}
    \maketitle

    \begin{abstract}
We have observed the B\,{\sc i} line at 2089.6\,{\AA}
in the metal poor star HD\,140283 with HST/STIS.
The observation was an attempt to confirm the B abundances derived from the
2497\,\AA\ B\,{\sc i} line in earlier works in general, and for for this
star in particular (Edvardsson et al. 1994;  Kiselman \& Carlsson 1996).
The resulting spectrum gained from 8 orbits of observations is hardly
consistent with the boron abundance derived in the earlier works.
A pure $^{10}$B line could produce the feature observed but no reasonable
process can produce such a ratio between $^{10}$B and $^{11}$B.
More likely the analysed feature is affected by a statistical fluctuation.
A conservative upper limit estimate of the NLTE corrected boron abundance
gives a value marginally consistent with that derived from the 2497\,\AA\
line.
We are proposing further observations to derive a more definite
abundance from the line.

    \end{abstract}

\section{Introduction}
The abundances of the lightest elements and their isotopes play a well-known
and important r\^ole in cosmology, for the early evolution of the Galaxy,
and for the understanding of stellar evolution.
Be and B abundances in the oldest stars have important implications for
the spallative interactions between cosmic rays and the ISM which are
thought to be mainly responsible for their production.
There is, however, still no generally accepted model for the details of
these mechanisms, and improved abundance determinations -- especially for
the extreme metal poor stars -- are very important
for the testing of different scenarios.
In conjunction with measurements of Li abundances, Be and B also probe
the convection zones and mixing processes in late-type stars.

In halo stars, all boron abundance determinations have so far been derived
from the 2496.7\,\AA\ B\,{\sc i} line.
This is an attempt to improve this state of affairs by observations of the
2089.6\,\AA\ B\,{\sc i} line in the well-known, bright halo dwarf star
HD\,140283 (BD\,-10 4149, HIP\,76976, SAO\,159459).
This line has the extra virtue of showing an unusually large isotope shift
between the two stable isotopes $^{11}$B and $^{10}$B,
(25\,m\AA, Johansson et al. 1993) which has already been used or tried for
isotope-ratio determinations in the ISM and in galactic disk- or thick-disk
stars.
This will eventually also be possible for halo dwarfs.

\section{Observations and data reductions}
The observations were obtained
with the Space Telescope Imaging Spectrograph (STIS)
onboard the Hubble Space Telescope (HST)
in February and April 1999 (proposal ID\,7348).
The STIS high-resolution echelle grating E230H, and the NUV-MAMA detector
were used, providing a spectral resolution of 110,000 per resolution 
element.
A total of 22 exposures of between 586 and 1120 seconds of length,
with three different wavelength settings (to reduce influence of non-uniform
detector sensitivity), and a total of 305 minutes exposure time were 
obtained
for the wavelength region 2083 - 2092\,\AA. In between the two visits
the detector had been shifted in the cross-dispersion direction
(this is done monthly), which provided a total
of six different positions on the detector for the line.

The STIS data reductions give calibrated spectra.
The transformation to a relative flux scale was performed by division
by a linear function, the slope of which was found by comparison with
a synthetic spectrum, i.e., no nocal continuum ``rectification'' has been done.
As seen in Fig.\,1, the fit to the continuum is quite satisfactory,
except possibly in the blue edge of the order.

The final S/N of the observed spectrum was estimated in the following
manner: the total time-weighted flux spectra from visit 1 and visit 2
were added, and also subtracted to get the difference.
The signal was measured in several different wavelength regions
of the added spectra.
The noise was measured in the same regions from the difference,
by calculating the (resolution element) pixel to pixel standard
deviation.
By using the difference between the spectra, the effect of undetected
weak lines on the noise are removed, leaving only the 'true'
noise.
The signal divided by the rms noise
gives our best estimate of the achieved S/N.
Note that different parts of the detector and calibration frames were
used in the two visits.

The final 2-pixel resolution-element S/N ratio obtained was $\sim 50$,
with a small variation between the different measurement sections.
This value is close to what the WWW STIS Exposure Time Calculator
presents, S/N $\sim 55$.
Part of the small difference could possibly be due to the shifting of
the spectra and pixel interpolation to a uniform wavelength scale.

The calibrations were initially done with the calibration
files available when the data was obtained, and later with
the latest calibration files available from May 2000.
No differences in the reduced spectra can be observed between the two
calibrations.

\section{Boron abundance analysis}
A standard LTE abundance analysis was performed using a MARCS model 
atmosphere
(Gustafsson et al., 1975; Asplund et al., 1997) with parameters
$T_{\rm eff}=5680$\,K, $\log g=3.5$, [Fe/H]\,$=-2.64$,
and $\xi_{\rm t}=1.5$\,km\,s$^{-1}$ adopted from Edvardsson et al. (1994).
The increased relative abundances of $\alpha$ elements in metal-poor stars
was accounted for by adopting [$\alpha/{\rm Fe]}=+0.4$, where 
even-atomic-number
elements from C to Ti were considered to be $\alpha$ elements.

Basic atomic line data (wavelengths, excitation energies, oscillator
strengths, radiation damping parameters and energy level designations)
for atomic and singly ionized lines in the wavelength region
2083 - 2092\,\AA\ were obtained from the VALD data base
(Piskunov et al. 1995; Ryabchikova et al. 1999; Kupka et al. 1999;
and a large number of references therein).
The treatment and parameters for ``van der Waals'' broadening was for most
strong lines obtained from the publications of O'Mara and colleagues,
see Barklem \& O'Mara (1998, and references therein).
For the remaining lines correction factors were applied to the classically
derived $\Gamma_6$ parameters as detailed in Edvardsson et al. (1993).
The line oscillator strengths were then modified to give a good general fit
of the line absorption in the wavelength region.

Simultaneously with the line data fitting, the absolute wavelength scale of
the observed spectrum was determined by fitting to the synthetic spectrum.
The final agreement for most of the strong lines is very good, and 
deficiencies
in the line list must be blamed for less god fits.
We estimate that the resulting absolute wavelength scale is better than
$\pm 5$\,m\AA. 
We have not fitted all lines rigorously, e.g. the line at 2091.7 \AA. 
Fig.\,1 shows the fit of the wavelength-adjusted observed spectrum to the
synthetic spectrum.

   \begin{figure*}
     \resizebox{\hsize}{!}{\includegraphics{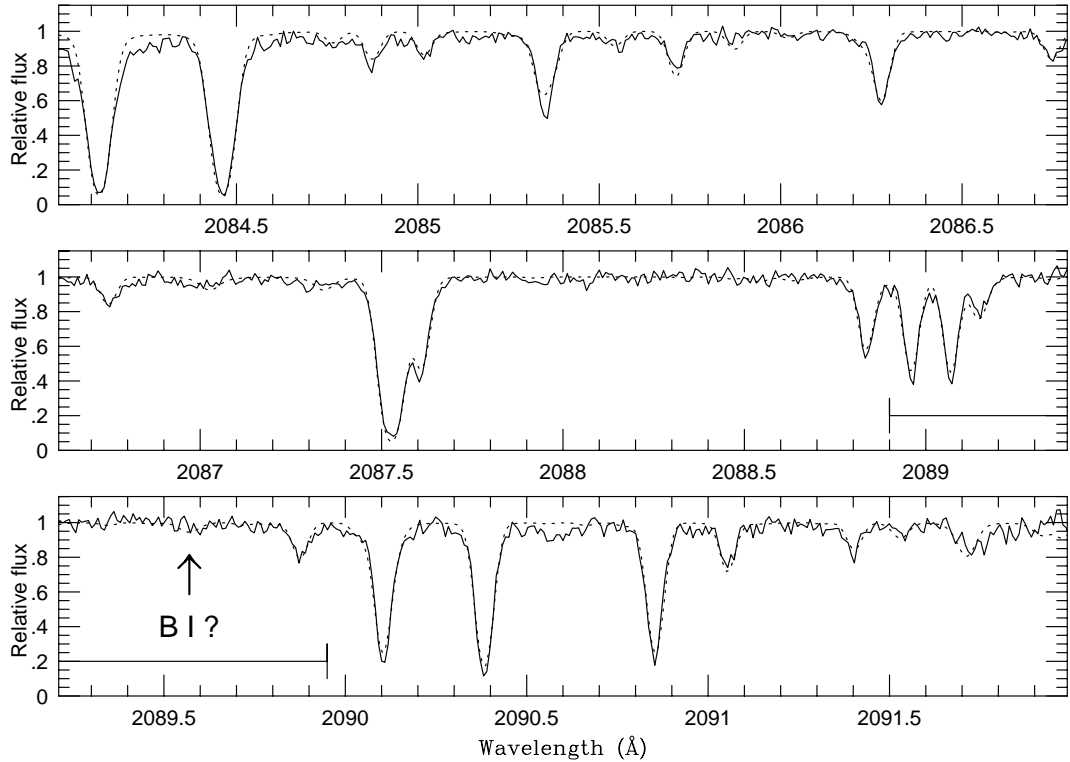}}
     \caption
     {The observed unbinned spectrum of HD\,140283 in the 2089.6\,\AA\ region
      ({\it solid line}).
      The wavelength scale of the observed spectrum was adjusted to fit the
      synthetic spectrum with adjusted $gf$ values ({\it dotted line}).
      The continuum level was fit by a single sloping line. Note that
      a higher continuum would ruin the fit in the 
      other parts of the spectrum
     }
   \end{figure*}

Wavelengths and oscillator strengths for B\,{\sc i} lines, including
isotopic shifts, were adopted from Johansson et al. (1993).
As an exercise and check we repeated the LTE analysis of the 2496.7\,\AA\
B\,{\sc i} line from Edvardsson et al. (1994), with identical result:
$\log (N_{\rm B}/N_{\rm H}) + 12.00 = -0.20$ (LTE).
Corrections for the simplifying but erroneous assumption of LTE
for boron for both the 2496.7\,\AA\ line (+0.52\,dex) and the 2089.6\,\AA\ 
line (+0.61\,dex) were adopted from
Kiselman \& Carlsson (1996, Table 3).
Synthetic spectra with $\log (N_{\rm B}/N_{\rm H}) + 12.00 = -0.29$ (LTE)
are compared with the observations in Fig.\,2.
This corresponds to a corrected abundance of
$\log (N_{\rm B}/N_{\rm H}) + 12.00 = +0.32$ (NLTE).
It should be noted that the differential NLTE correction of 0.09\,dex
between the two boron lines is quite insensitive to uncertainties in the
NLTE analysis (Kiselman, 2000).

   \begin{figure*}
     \resizebox{\hsize}{!}{\includegraphics{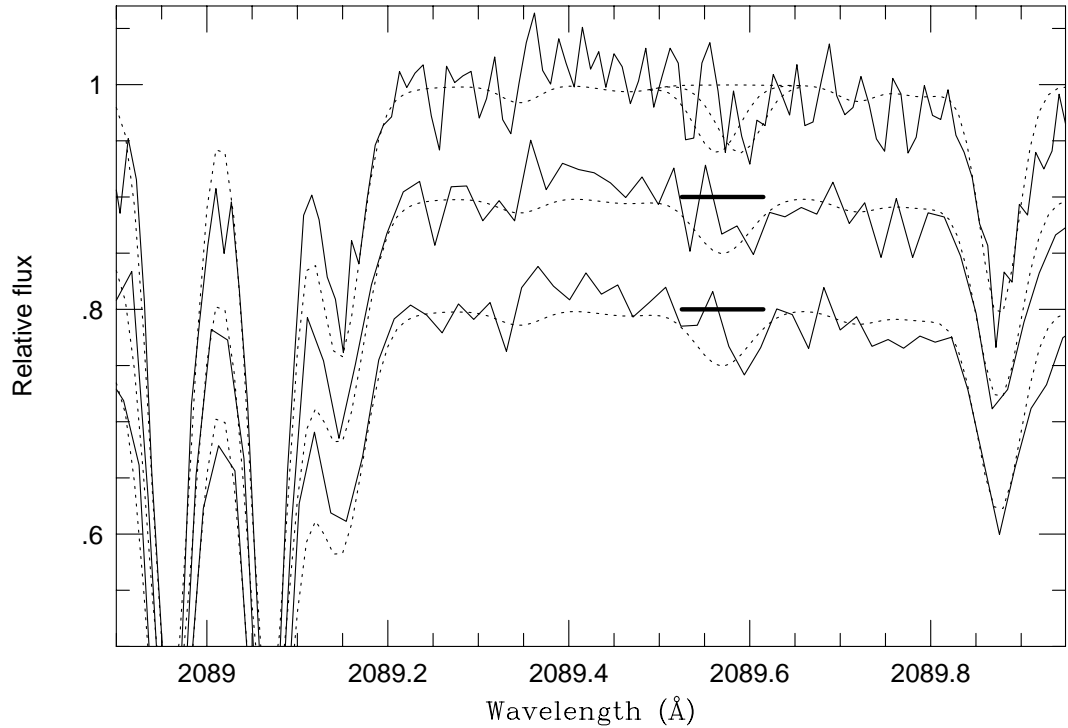}}
     \caption
     {Top: The observed spectrum of HD\,140283 in the 2089.6\,\AA\ region
      ({\it solid line}).
      The three {\it dotted lines} show our synthetic spectra, one without
      any boron, and the two other ones with the B\,{\sc i}
      abundance adopted from the 2497\,\AA\ line by
      Edvardsson et al. (1994): the bluer one with
      only $^{11}$B, and the more red one with only $^{10}$B.
      (The NLTE corrections of Kiselman \& Carlsson (1996) are
      consistently taken into account.)
      The two lower spectra show the two possible two-pixel binned versions
      corresponding to the nominal resolution of the observations and a
      synthetic spectrum with $^{11}{\rm B}/^{10}{\rm B}=2.5$ as expected
      from high-energy spallation reactions.
      We can only determine an upper limit for the boron abundance and are
      proposing further observations to ascertain detection of the boron line
     }
   \end{figure*}

As seen in Fig.\,2 we can not claim a detection of the boron feature.
One might suspect a weak absorption near 2089.60\,\AA\ which would be best
fit by a pure $^{10}$B line (the 25\,m\AA\ isotope shift is indicated at
the top of Fig.\,2).
We do not know, however, of any process which would be expected to produce
predominantly $^{10}$B, and we consider the unexpected line position to be
most probably due to a statistical fluctuation.
We try instead to estimate an upper limit for the boron abundance:
The width over which the expected boron feature should mainly depress the
continuum is about 90\,m\AA\ or 5 resolution elements (10 pixels).
This region is indicated by the thick solid lines in Fig.\,2.
The integrated absorption equivalent width in this interval of the
un-binned observed spectrum is 2.3\,m\AA.
With a $S/N = 50$ per resolution element, we have $S/N = 50 \sqrt{5} = 112$
for the line, which gives a $1 \sigma$ equivalent width uncertainty of
$90/112=0.8$\,m\AA.
{}From the discussion of the continuum level determination procedure and
Figs.\,1 and 2, we judge that the continuum level may at the most
be wrong by 1.5\%, which would correspond to 1.4\,m\AA\ to the line.
Adding these two uncertainties in quadrature makes an estimated $1 \sigma$
equivalent width uncertainty of 1.6\,m\AA.
{}From these considerations we estimate that a $1 \sigma$
upper limit of the equivalent width is $2.3 + 1.6 = 3.9$\,m\AA.
This value is just barely consistent with the abundance
derived from the 2496.7\,\AA\ line by Edvardsson et al. (1994).
Note, however, that the center-of-gravity of the possibly present feature
is shifted by 25\,m\AA\ from where it would be expected.

A simple Monte-Carlo simulation with 500 realizations of Gaussian noise added
to a synthetic spectrum confirms the equivalent width uncertainty estimate
above.
It also indicates that a $1 \sigma$ wavelength shift due to noise should
be 30\,m\AA.

In conclusion:
The possible feature is about $1 \sigma$ weaker 
and a little less than $1 \sigma$ shifted in wavelength compared to the
{\it bona fide} expected 4.0\,m\AA \ equivalent width and position.

\section{Systematic errors}
Since our worries are mainly due to the different results from the two
boron lines, we especially have to care about errors which may affect the
strengths of the two weak boron lines differently.

The effective temperature scale for late-type stars is continuously
under debate.
For this particular star estimates differ by about 200\,K.
We find that even with a $\pm$200\,K $T_{\rm eff}$ variation, the relative
strength of the lines should not vary by more than 2\%.

The continuous opacities in the UV region of cool stars are still
under discussion.
There are suggestions that we are lacking important opacity
in e.g. the Sun (e.g Balachandran \& Bell, 1998).
The ``missing opacity'' is, however, generally ascribed to metals and
therefore of importance mainly in metal-rich stars.
HD\,140283 is an extreme Pop II star (the metallicity is about 1/400 of
that of the Sun), and uncertainties in the continuous metal opacity
are unimportant.

Both lines emanate from the B\,{\sc i} ground state, they are
formed at similar optical depths, and have similar NLTE corrections.
The possible errors in the NLTE calculations should also be similar for
the two.
Therefore such errors would have a small effect on this
differential analysis.
The relative NLTE effects are also quite insensitive to variations
in $T_{\rm eff}$ for these model parameters
(see Fig. 9 of Kiselman \& Carlsson, 1996).

Any unknown line blending with the 2089.6\,\AA\ line would only worsen our
``problem'' and further decrease the boron abundance.
A blend with the 2497\,\AA\ line on the other hand might explain our
result.
The effect of this would probably be that all stellar boron abundance
determinations would have to be revised.

\section{The line in HD\,76932}
The 2089.6\,\AA\ line was observed in the [Fe/H]$= -1.0$
thick-disk star HD\,76932 by Rebull et al. (1998).
The 2089.6\,\AA\ feature is strong in that star, actually stronger than
expected from the 2497\,\AA\ line, even when NLTE corrections are not
taken into account.
The line position furthermore tends to suggest a $^{11}$B-rich feature.
Both these findings are opposite to the ones we find for our
extreme Pop\,II star.
Rebull et al. suggest that there may be a blending line affecting the
2089.6\,\AA\ feature in HD\,76932.
If that is so, we see no traces of it in HD\,140283.

\section{Conclusions}
We have used eight HST/STIS orbits to observe the 2089.6\,\AA\ B\,{\sc i}
line in HD\,140283.
The observed spectrum is only marginally consistent with the boron abundance
found in previous determinations based on the 2496.7\,\AA\ B\,{\sc i} line.
We are proposing further HST/STIS observations to secure the detection of
the 2089.6\,\AA\ B\,{\sc i} line, as a very important check on all previous
boron abundance determinations for halo stars.

\begin{acknowledgements}
We thank Dr. Dan Kiselman for valuable discussions on boron-line formation and
NLTE effects and also the referee Dr. Werner W. Weiss for valuable comments to the manuscript.
Professor Bengt Gustafsson is thanked for helpful and enlightening discussions.
The authors are supported by the Swedish Natural Sciences Research Council
and the Swedish National Space Board.
\end{acknowledgements}

\end{document}